\documentclass[twocolumn,prb]{revtex4}

\usepackage{amsfonts}
\usepackage[T1]{fontenc}
\usepackage{amsmath,amsbsy,amssymb,graphicx}
\usepackage{times}

\begin{document}

\title{Nonlinear Anderson localization in Toda lattices}
\author{Motohiko Ezawa}
\affiliation{Department of Applied Physics, University of Tokyo, Hongo 7-3-1, 113-8656,
Japan}

\begin{abstract}
We study the Anderson localization in nonlinear systems by taking a
nonlinear transmission line realizing the Toda lattice. It is found that the
randomness in inductance induces the Anderson localization in the voltage
propagation. Furthermore, the nonlinearity enhances the Anderson
localization. They are understood in terms of attractive interactions to
form a soliton in a nonlinear system. Our results will be applicable in
general to the Anderson localization in nonlinear systems that have solitons.
\end{abstract}

\maketitle

\section{Introduction}

The Anderson localization is a prominent phenomenon inherent to disordered
systems\cite{Anderson}, where the system exhibits an insulating behavior due
to disorders. It is caused by the interference of wave functions due to
multiple-scattering paths under a random potential. In the strong scattering
limit, the wave functions are completely trapped by a disordered medium. The
Anderson localization has been discussed in various systems\cite%
{Cutler,Gia,Roati,Segev} including nonlinear systems\cite{Piko,Lahini}. It
is pointed out that the Anderson localization is destructed due to a
nonlinearity in the case of the discrete Anderson nonlinear Schr\"{o}dinger
equation\cite{Piko}. On the other hand, the enhancement and delocalization
of the Anderson localization are reported in photonic lattices\cite{Lahini}.
It is an intriguing problem to make a further investigation how the
nonlinearity affects the Anderson localization.

Electric circuits attract renewed interest in the context of topological
physics\cite%
{TECNature,ComPhys,Hel,Lu,YLi,EzawaTEC,Research,Zhao,EzawaLCR,EzawaSkin,Garcia,Hofmann,EzawaMajo,Tjunc,Lee}
based on the observation that a circuit Laplacian may be identified with a
Hamiltonian. Furthermore, the telegrapher equation is rewritten in the form
of the Schr\"{o}dinger equation, and thus describes the behavior of a
quantum walker\cite{QWalk,EzawaUniv,EzawaDirac}. One of the merits of
electric circuits is that we can naturally introduce nonlinear elements into
circuits. Hence, electric circuits present an ideal playground to study
nonlinear physics. There are also some studies on topological physics in
nonlinear electric circuits\cite{Kot,Sone,TopoToda}.

The Toda lattice is a celebrated nonlinear exactly solvable model possessing
soliton solutions\cite{Toda,Toda2,Toda3}. It is realized in a nonlinear
transmission line consisting of variable capacitance diodes and inductors%
\cite{Hirota,Singer,Yemele,Yemele2,Pelap05,Houwe,Nakajima}, as illustrated
in Fig.\ref{FigIllust}(a). Recently, it is shown that the dimerized Toda
lattice has a topological phase\cite{TopoToda} which is a nonlinear
extension of the Su-Schrieffer-Heeger model.

In this paper, we study an interplay between the Anderson localization and
nonlinear effects taking an instance of a nonlinear transmission line
realizing the Toda lattice. We introduce randomness into the inductance of
the transmission line and investigate the dynamics of the voltage
propagation starting from one node. Namely, we analyze a propagation
dynamics of the voltage by giving a delta-function type pulse at one node.
The diffusion process strongly depends on the strength of the disorder. It
is found that the voltage propagation shows a localization behavior when
disorders are strong enough, which is the nonlinear Anderson localization in
the nonlinear transmission line. Furthermore, it is found that the
nonlinearity enhances the Anderson localization.

We interpret these phenomena as caused by the presence of the attractive
nonlinear force in a nonlinear system which has a soliton. Indeed, in a
soliton system, there is an attractive force in order to prevent a soliton
from decaying. Without the attractive force, the waves with the different
momenta propagate with different velocities, resulting in the destruction of
a soliton. This attractive force maintains the wave packet starting from the
delta function even though it is not a soliton solution. Moreover, once one
of the wave components is trapped to a randomness, the other components are
also trapped, which enhances the localization in the nonlinear system. Then,
it is natural that the localization is strong when the nonlinearity is
strong. Our results will be applicable in general to the nonlinear Anderson
localization in nonlinear systems having solitons.

\section{Toda lattice}

\begin{figure}[t]
\centerline{\includegraphics[width=0.48\textwidth]{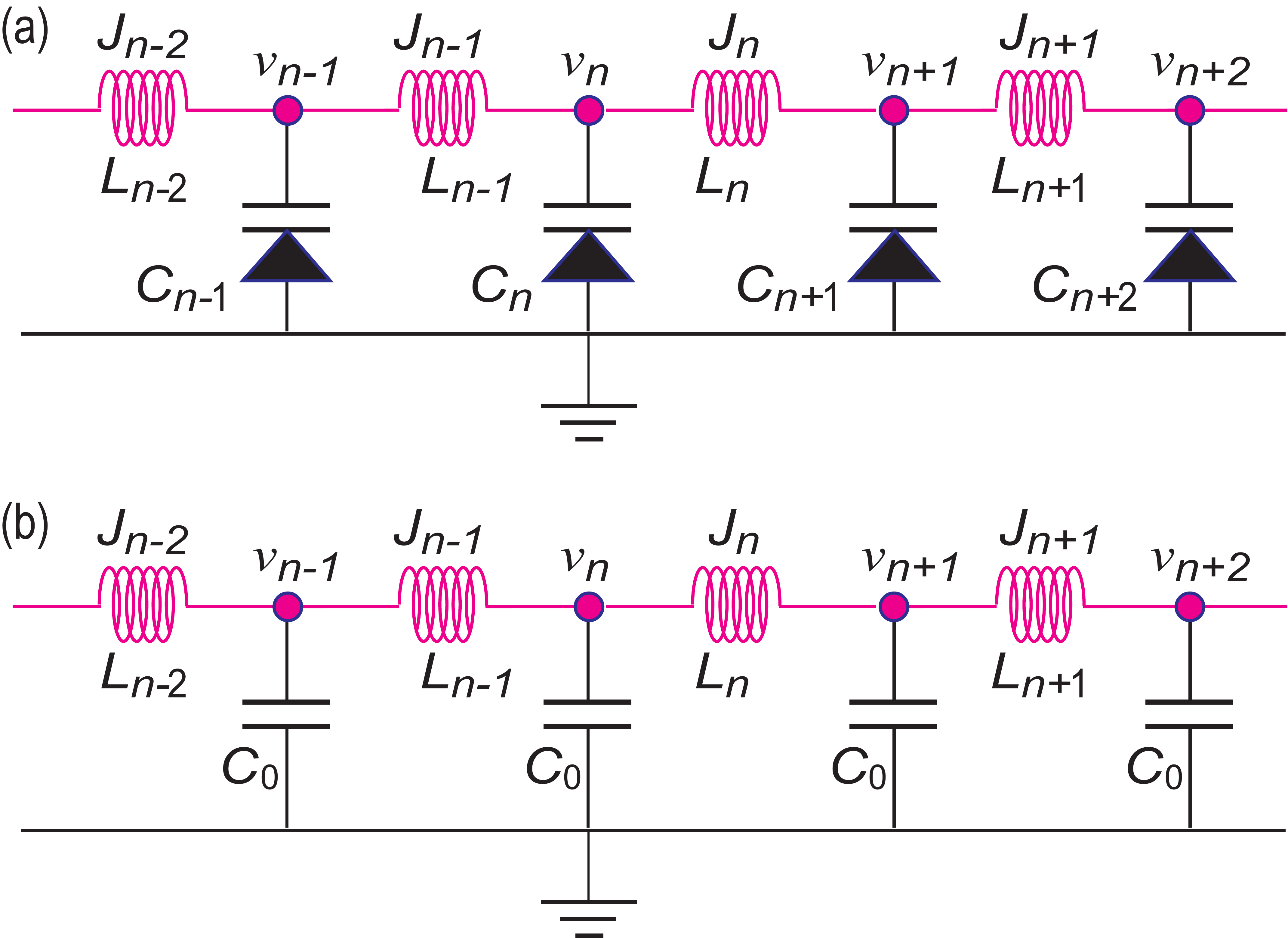}}
\caption{Illustration of a transmission line made of (a) nonlinear elements
realizing the inhomogeneous Toda lattice and (b) linear elements realizing
the telegrapher equation. The inductance indicated by magenta is random.
Each node is grounded via a variable-capacitance diode and
a capacitor in (a) and (b), respectively.}
\label{FigIllust}
\end{figure}

We review the Toda model. The Toda Hamiltonian is defined by\cite{Toda,Toda2}%
\begin{equation}
H=\sum_{n}\frac{m}{2}\left( \frac{dy_{n}}{dt}\right) ^{2}+\Phi \left(
r_{n}\right) ,  \label{TodaH}
\end{equation}%
where $r_{n}=y_{n+1}-y_{n}$, and $\Phi \left( r_{n}\right) $ is the Toda
potential, 
\begin{equation}
\Phi \left( r_{n}\right) \equiv \frac{a}{b}e^{-br_{n}}+ar_{n}-\frac{a}{b}.
\end{equation}%
A nonlinear differential equation follows from Eq.(\ref{TodaH}),%
\begin{equation}
m\frac{d^{2}y_{n}}{dt^{2}}=f_{n-1}-f_{n},  \label{TodaA}
\end{equation}%
where $f_{n}$ is the force,%
\begin{equation}
f_{n}=-\frac{d}{dr_{n}}\Phi \left( r_{n}\right) .
\end{equation}%
The Toda equation (\ref{TodaA}) is rewritten in the form of%
\begin{equation}
m\frac{d^{2}}{dt^{2}}\log \left( 1+\frac{f_{n}}{a}\right) =b\left(
f_{n+1}+f_{n-1}-2f_{n}\right) ,  \label{TodaEq}
\end{equation}%
which is well realized by a transmission line with variable-capacitance
diodes\cite{Hirota}: See Fig.\ref{FigIllust}(a) and Eq.(\ref{TeleToda}).

The right hand side of Eq.(\ref{TodaEq}) is identical to the tight-binding
hopping Hamiltonian between the nearest neighbor sites,%
\begin{equation}
H=\sum_{n}M_{nm}\left\vert \psi _{n}\right\rangle \left\langle \psi
_{m}\right\vert ,  \label{Hop}
\end{equation}%
where $M_{nm}=b\left( \delta _{n-1,m}+\delta _{n+1,m}-2\delta _{nm}\right) $
with $b$ the hopping parameter.

\section{Inhomogeneous Toda Lattice}

We generalize the Toda lattice to an inhomogeneous Toda lattice\cite%
{TopoToda}, because the hopping becomes node-dependent in the presence of
randomness in the transmission line. We show the transmission line
corresponding to the inhomogeneous Toda lattice in Fig.\ref{FigIllust}(a),
which consists of variable-capacitance diodes and inductors. The Kirchhoff
law is given by 
\begin{align}
L_{n}\frac{dJ_{n}}{dt}=& v_{n}-v_{n+1}, \\
\frac{dQ_{n}}{dt}=& J_{n-1}-J_{n},
\end{align}%
where $v_{n}$ is the voltage, $J_{n}$ is the current and $Q_{n}$ is the
charge at the node $n$, while $L_{n}$ is the inductance for the inductor
between the nodes $n$ and $n+1$, as illustrated in Fig.\ref{FigIllust}(a).
The Kirchhoff law is summarized in the form of the second-order differential
equation, 
\begin{align}
\frac{d^{2}Q_{n}}{dt^{2}}=& \frac{dJ_{n-1}}{dt}-\frac{dJ_{n}}{dt}  \notag \\
=& \frac{1}{L_{n-1}}\left( V_{n-1}-V_{n}\right) -\frac{1}{L_{n}}\left(
V_{n}-V_{n+1}\right) ,  \label{EqC}
\end{align}%
where we have introduced a new variable $V_{n}$ by $v_{n}=V_{0}+V_{n}$.

The capacitance is a function of the voltage $V_{n}$\ in the
variable-capacitance diode, and it is well given by\cite{Nakajima}%
\begin{equation}
C\left( V_{n}\right) =\frac{Q\left( V_{0}\right) }{F_{0}+V_{n}-V_{0}},
\label{CV}
\end{equation}%
where $F_{0}$ is a constant characteristic to the variable-capacitance
diode. Especially, we have%
\begin{equation}
C\left( V_{0}\right) =\frac{Q\left( V_{0}\right) }{F_{0}}.
\end{equation}%
The charge is given by%
\begin{align}
Q_{n}=& \int_{0}^{V_{n}}C\left( V\right) dV  \notag \\
=& Q\left( V_{0}\right) \log \left[ 1+\frac{V_{n}}{F_{0}}\right] +\text{%
const.}  \label{EqD}
\end{align}%
A closed form of the differential equation for $V_{n}$ follows from Eqs.(\ref%
{EqC}) and (\ref{EqD}), 
\begin{align}
Q& \left( V_{0}\right) \frac{d^{2}}{dt^{2}}\log \left[ 1+\xi \frac{V_{n}}{%
V_{1}}\right]  \notag \\
& =\frac{V_{n+1}}{L_{n}}-\left( \frac{1}{L_{n-1}}+\frac{1}{L_{n}}\right)
V_{n}+\frac{V_{n-1}}{L_{n-1}},  \label{EqA}
\end{align}%
where we have introduced a dimensionless\ nonlinearity parameter 
\begin{equation}
\xi \equiv V_{1}/F_{0}.
\end{equation}
Small $\xi $ implies weak nonlinearity, while large $\xi $ implies strong
nonlinearity. It is an inhomogeneous generalization of the Toda equation.

\begin{figure}[t]
\centerline{\includegraphics[width=0.48\textwidth]{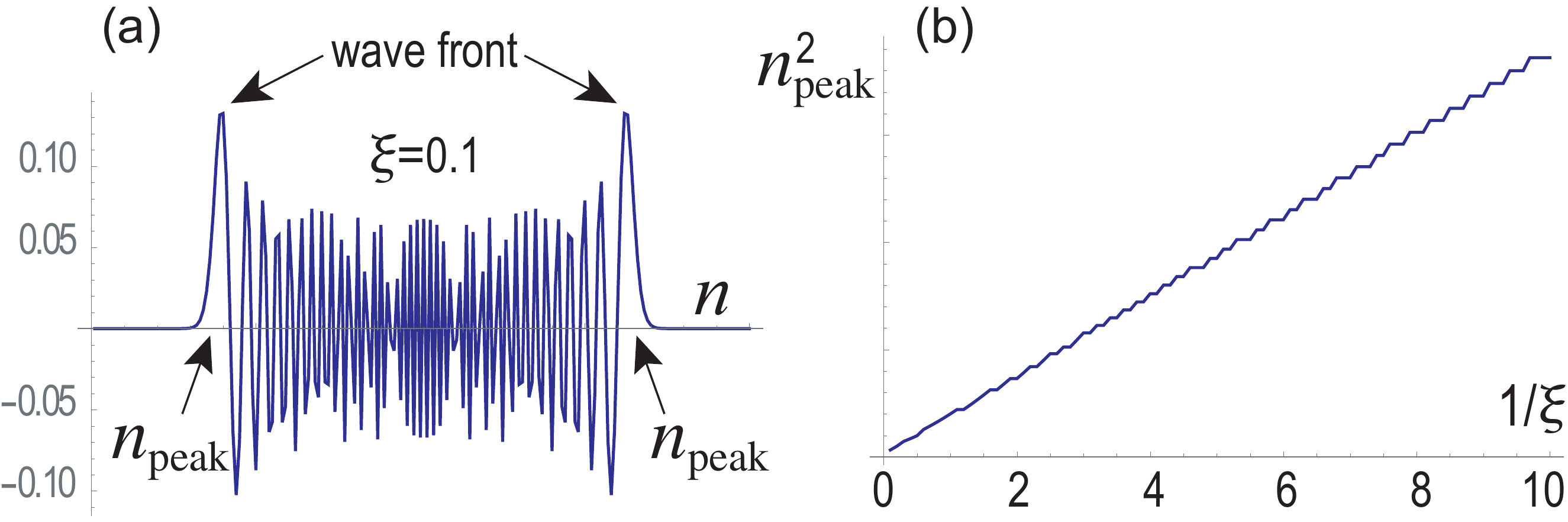}}
\caption{(a) Wave pattern of the voltage $V_{n}$ at a fixed time $t$ with $%
\protect\xi =0.1$.\ The vertical axis is the voltage in the unit of $V_{1%
\text{. }}$(b) $n_{\text{peak}}^{2}$ as a function of the nonlinearity $%
\protect\xi $ at a fixed time $t$. We have set $Q\left( V_{0}\right) =L=1$.}
\label{FigFront}
\end{figure}

We first study the homogeneous case $L_{n}=L$. Eq.(\ref{EqA}) is reduced to%
\begin{equation}
Q\left( V_{0}\right) \frac{d^{2}}{dt^{2}}\log \left[ 1+\xi \frac{V_{n}}{V_{1}%
}\right] =\frac{1}{L}\left( V_{n+1}-2V_{n}+V_{n-1}\right) ,  \label{TeleToda}
\end{equation}%
which is the Toda equation (\ref{TodaEq}) with $m=Q\left( V_{0}\right) $, $%
a=F_{0}$ and $b=1/L$. We set an initial condition, 
\begin{equation}
V_{n}\left( 0\right) =V_{1}\delta _{n,n_{1}},  \label{InitialCon}
\end{equation}%
where the voltage is nonzero only at the node $n_{1}$.

\begin{figure}[t]
\centerline{\includegraphics[width=0.48\textwidth]{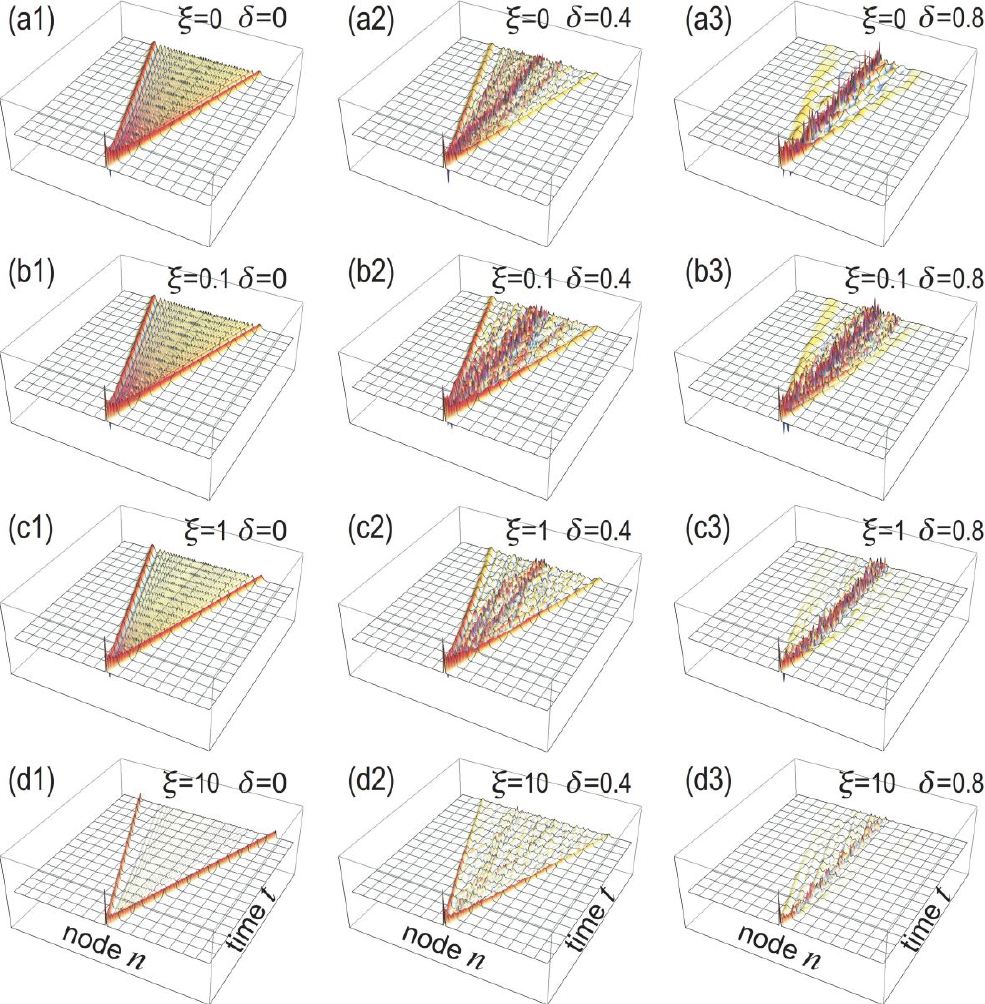}}
\caption{Dynamics of the voltage propagation. (a1)$\sim $(d1) $\protect%
\delta =0$, (a2)$\sim $(d2) $\protect\delta =0.4$, and (a3)$\sim $(d3) $%
\protect\delta =0.8$. (a1)$\sim $(a3) Linear wave propagation corresponding
to $\protect\xi =0$. We have set (b1)$\sim $(b3) $\protect\xi =0.1$, (c1)$%
\sim $(c3) $\protect\xi =1$ and (d1)$\sim $(d3) $\protect\xi =10$. The
vertical axis is the dimensionless voltage $x_{n}$. }
\label{FigDiffuse}
\end{figure}

The Toda equation (\ref{TeleToda}) is reduced to a linear wave equation by
taking the limit $\xi \rightarrow 0$ in (\ref{CV}),%
\begin{equation}
C\left( V_{0}\right) \frac{d^{2}V_{n}}{dt^{2}}=\frac{1}{L}\left(
V_{n+1}-2V_{n}+V_{n-1}\right) .  \label{EqB}
\end{equation}%
Correspondingly, a nonlinear transmission line [Fig.\ref{FigIllust}(a)] is
reduced to a linear transmission line [Fig.\ref{FigIllust}(b)], which we
have already studied in the context of quantum walks\cite{QWalk}.

\section{Dynamics of the voltage propagation}

We investigate the time evolution of the voltage $V_{n}$\ by solving the
Toda equation (\ref{TeleToda}) numerically together with the initial
condition (\ref{InitialCon}). In Fig.\ref{FigFront}(a), we show the wave
pattern of $V_{n}$\ as a function of the node $n$ at a fixed time, where the
absolute value of the voltage takes the largest value at the wave front
whose position is indicated by $n_{\text{peak}}$. In Fig.\ref{FigDiffuse}%
(a1), (b1), (c1) and (d1), we show the time evolution of the voltage
propagation for $\xi =0$, $0.1$, $1$ and $10$, respectively. It shows a
diffusive motion of a wave, where the wave front is indicated in red. It is
found to be a linear function of the time $t$, $n_{\text{peak}}\varpropto t$%
. Next, we\ have plotted the peak position $n_{\text{peak}}^{2}$ as a
function of $\xi ^{-1}$ with all other variables fixed in Fig.\ref{FigFront}%
(b). It is clearly a linear function of $\xi ^{-1}$, \ or $n_{\text{peak}%
}\varpropto \xi ^{-1/2}$. Consequently, it follows that%
\begin{equation}
n_{\text{peak}}\varpropto \xi ^{-1/2}t\varpropto \sqrt{F_{0}}t,
\label{WaveFront}
\end{equation}%
when $L$ and $Q(L_{0})$ are fixed.

We note that the ripples are confined between two wave fronts and absent
outside of the wave fronts. This resembles the light cone, where the
information propagates with a finite velocity.

A prominent feature of the wave pattern reads as follows: There are many
ripples between two wave fronts in Fig.\ref{FigDiffuse}(a1) and (b1). In
particular, Fig.\ref{FigDiffuse}(a1) is for the voltage propagation in the
linear limit, where the delta-function like pulse disperses quickly because
its Fourier components propagate freely. This is approximately the case
provided the nonlinearity is weak as in Fig.\ref{FigDiffuse}(b1) or in Fig.%
\ref{FigFront}(a). On the other hand, the ripples are weak and the two wave
fronts look as if they are two isolated solitary waves in strong nonlinear
systems as in Fig.\ref{FigDiffuse}(d1).

\section{Toda soliton solution}

It is well known that the Toda equation (\ref{TeleToda}) has an exact
soliton solution. By assuming the Toda soliton starts from $n=1$ at $t=0$,
it is given by\cite{Toda,Toda2,Toda3}%
\begin{equation}
V_{n}=F_{0}\sinh ^{2}\kappa \,\text{sech}^{2}\left( \kappa n-\beta t\right) ,
\label{TodaSoliton}
\end{equation}%
where 
\begin{equation}
\beta =\pm \sqrt{\frac{F_{0}}{LQ\left( V_{0}\right) }}\sinh \kappa =\pm 
\frac{1}{\sqrt{LC\left( V_{0}\right) }}\sinh \kappa ,
\end{equation}%
and $\kappa $ is a constant determining the width of the soliton. In Fig.\ref%
{FigSoliton}, we show the time evolution of the voltage $V_{n}$\ subject to
the Toda equation (\ref{TeleToda}) for various $\kappa $. The soliton splits
into two solitons and propagate in two directions without tails.

It follows from Eq.(\ref{TodaSoliton}) that the position $n_{\text{Toda}}$
of the Toda soliton moves as%
\begin{equation}
n_{\text{Toda}}=\frac{\beta }{\kappa }t=\pm \sqrt{\frac{1}{LQ\left(
V_{0}\right) }}\frac{\sinh \kappa }{\kappa }\sqrt{F_{0}}t,
\end{equation}%
which is consistent with Eq.(\ref{WaveFront}).

\begin{figure}[t]
\centerline{\includegraphics[width=0.48\textwidth]{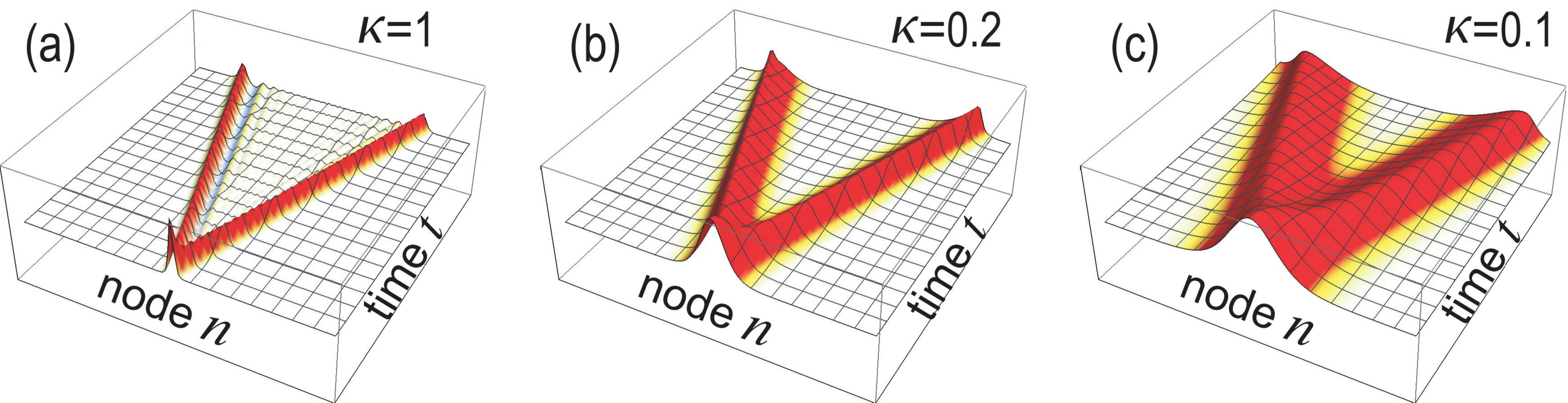}}
\caption{Propagation dynamics of the Toda soliton. (a) $\protect\kappa =1$,
(b) $\protect\kappa =0.2$ and (c) $\protect\kappa =0.1$. The vertical axis
is the voltage $V_{n}$. We have set $C(V_{0})=0.1$ and $L=F_{0}=1$.}
\label{FigSoliton}
\end{figure}

Let us make a qualitative argument with respect to the propagations of the
wave front and the Toda soliton. First, it is found that the propagation of
the Toda soliton for $\kappa =1$ in Fig.\ref{FigSoliton}(a) is quite similar
to that of the wave front for $\xi =10$ in Fig.\ref{FigDiffuse}(d1). It
seems that the Toda soliton becomes the delta-function type as $\kappa
\rightarrow \infty $, while the wave front becomes the delta-function type
as $\xi \rightarrow \infty $.

\begin{figure*}[t]
\centerline{\includegraphics[width=0.98\textwidth]{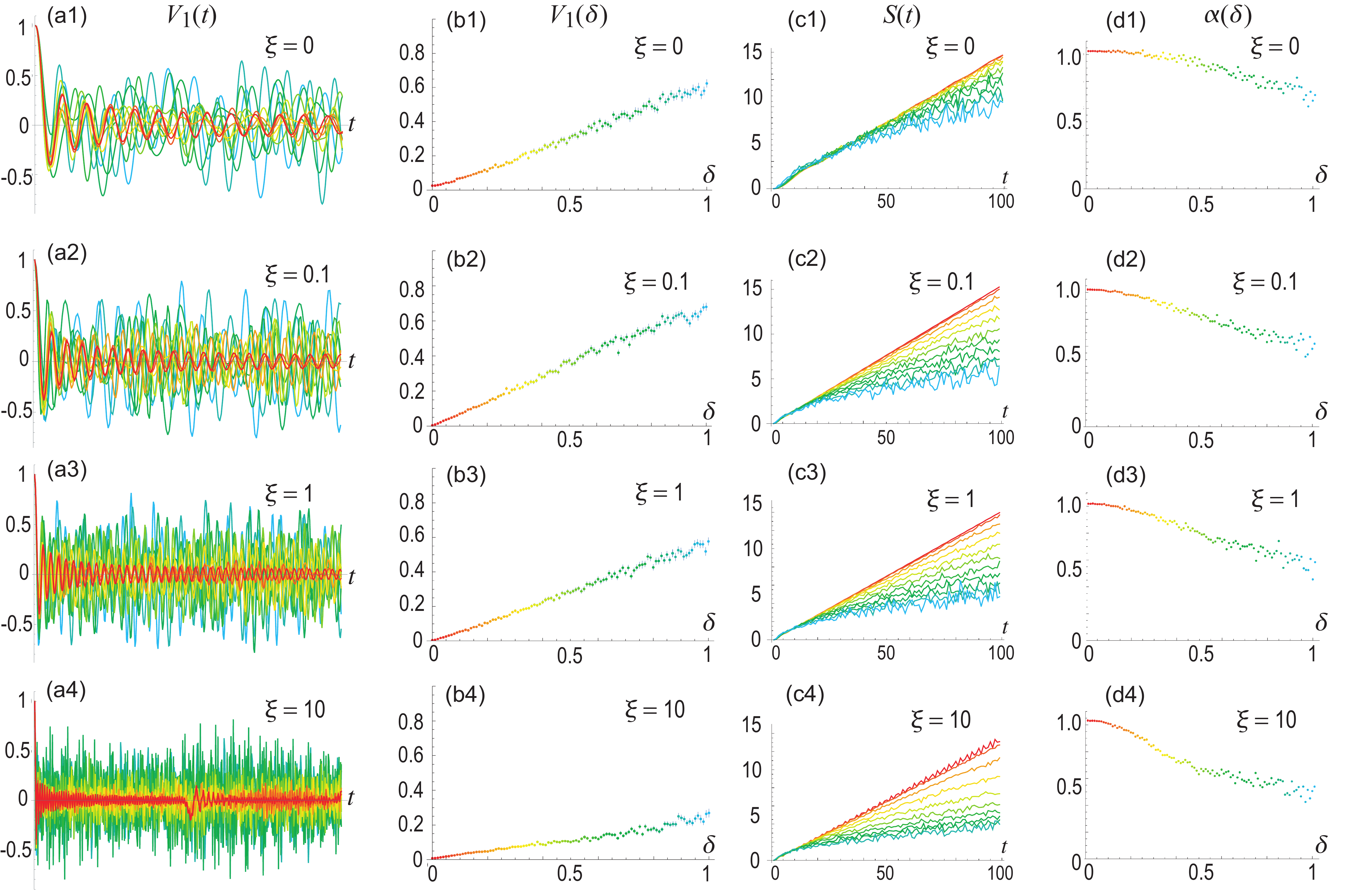}}
\caption{(a1)$\sim $(a4) Time evolution of the voltage at the initial node $%
n_{1}$. The vertical axis is the voltage in the unit of $V_{1}$. (b1)$%
\sim $(b4) The voltage at the initial node after enough time. Red color
indicates $\protect\delta =0$, while cyan color indicates $\protect\delta =1$%
. (c1)$\sim $(d4) 100 times average of time evolution of the standard
deviation $\ S\left( t\right) $ for various disorder strength $\protect%
\delta $ in the unit of $V_{1}$. 
(d1)$\sim $(d4) The fitted standard deviation
exponent $\protect\alpha $ as a function of disorder strength $\protect%
\delta $. (a1)$\sim $(d1) Linear wave ($\protect\xi =0$). We have set (a2)$%
\sim $(d2) $\protect\xi =0.1$, (a3)$\sim $(d3) $\protect\xi =1$ and (a4)$%
\sim $(d4) $\protect\xi =10$. We have set $C(V_{0})=0.1$ and $L=1$. }
\label{FigDeviate}
\end{figure*}

Indeed, when the soliton width is small enough, we may approximate Eq.(\ref%
{TodaSoliton}) as%
\begin{equation}
V_{n}=2F_{0}\sinh ^{2}\kappa \,\delta (\kappa n-\beta t).
\end{equation}%
The narrow soliton corresponds to the limit $\kappa \rightarrow \infty $,
which corresponds to the limit $\xi \rightarrow \infty $ to keep the voltage
finite.

We summarize our observation with respect to the voltage propagation in Fig.%
\ref{FigDiffuse}(b1)$\sim $(d1) as follows. In a soliton system, there is an
attractive force in order to form a soliton. Without the attractive force,
the waves with the different momenta propagate with different velocities,
resulting in the destruction of a soliton. This attractive force maintains
the wave packet starting from the delta function even though it is not a
soliton solution.

\section{Effects of randomness}

We introduce randomness into inductors uniformly distributing from $-\delta $
to $\delta $. As a result the inductance $L$\ becomes node-dependent, $%
L\rightarrow L_{n}$, where%
\begin{equation}
L_{n}=L\left( 1+\eta _{n}\delta \right) ,  \label{Ln}
\end{equation}%
with $\eta _{n}$ being a random variable ranging from $-1$ to $1$. Typical
values are $\delta =0.05\sim 0.2$ in electric circuits.

We now solve the inhomogeneous Toda equation (\ref{EqA}) with the use of Eq.(%
\ref{Ln}). We show the time evolution of the voltage for various $\delta $
and $\xi $ in Fig.\ref{FigDiffuse}. We show the results with $\delta =0.4$
in Fig.\ref{FigDiffuse}(a2)$\sim $(d2), where there remains a localized
component at the initial node $n_{1}$ although the wave fronts are clearly
seen. The voltage is almost localized at the initial node $n_{1}$ for $%
\delta =0.8$, as shown in Fig.\ref{FigDiffuse}(a3)$\sim $(d3).

Next we show the time evolution of the voltage at the initial node $n_{1}$
for various $\xi $ in Fig.\ref{FigDeviate}(a1)$\sim $(a4). The voltage
rapidly decreases in the absence of the disorder , as shown by the red
curves ($\delta =0$). The voltage can be checked to vanish after enough
time, although such a limit is not displayed in these figures. On the other
hand, the voltage remains nonzero when the disorder is introduced as
indicated by cyan curves ($\delta =1$). We plot the voltage after enough
time in Fig.\ref{FigDeviate}(b1)$\sim $(b4). The remnant voltage is almost
proportional to the disorder strength $\delta $. We note that there is no
significant dependence on $\xi $.

We define the standard deviation of the voltage propagation by%
\begin{equation}
S\left( t\right) \equiv \sqrt{\sum_n \left( n-n_{1}\right) ^{2}V_{n}^{2}\left(
t\right) },
\end{equation}%
where $n_{1}$ is the initial node. We show the time evolution of the
standard deviation in Fig.\ref{FigDeviate}(c1)$\sim $(c4). It increases
almost linearly as a function of the time. It indicates a diffusion process
of the quantum-walk type. We approximate $S\left( t\right) $ by $S\left(
t\right) \propto t^{\alpha }$ as a function of $t$. We show the exponent $\alpha $ 
in Fig.\ref{FigDeviate}(d1)$\sim $(d4). It is almost constant $\alpha =1$ 
for $\delta <0.4$ for $\xi =0$ and $\delta <0.1$ for $\xi=0.1,1,10$, 
while it decreases gradually as the increase of $\delta $. The
exponent $\alpha$ decreases rapidly to $\alpha \simeq 0.5$ for strong
nonlinearity $\xi $, as shown in Fig.\ref{FigDeviate}(d4). It indicates that
the nonlinearity enhances the Anderson localization.

\section{Discussion}

We discuss why the Anderson localization enhances in the nonlinear system
with the Toda soliton. In the linear wave equation, the waves with the
different momenta propagate independently. On the other hand, the waves do
not propagate independently in nonlinear wave equations. Especially, there
is such an interaction in a soliton system that prohibits the soliton from
decaying. Hence, there must be a strong attractive force preventing a
solitary wave from diffusing. We interpret this phenomenon as caused by the
presence of the attractive nonlinear force forming a soliton in the Toda
model. Once one of the wave components is trapped to a randomness, the other
components are also trapped, which enhances the localization in the
nonlinear system. Furthermore, the localization is stronger when the
nonlinearity is stronger. Our results will be applicable to the nonlinear
Anderson localization in other soliton systems. It is interesting to study
the Anderson localization in other nonlinear systems.

In conclusion, we have studied nonlinear Anderson localization in Toda
lattices, where we find that the nonlinearity enhances the localization. Our
results will excite further studies on the relevance of the nonlinearity to
the Anderson localization.

The author is very much grateful to M. Kawamura, S. Katsumoto and N. Nagaosa
for helpful discussions on the subject. This work is supported by the
Grants-in-Aid for Scientific Research from MEXT KAKENHI (Grants No.
JP17K05490 and No. JP18H03676). This work is also supported by CREST, JST
(JPMJCR16F1 and JPMJCR20T2).

\end{document}